\documentclass[
reprint,
 amsmath,amssymb,
aps,
prb,
floatfix
]{revtex4-2}

\usepackage{graphicx}
\usepackage{dcolumn}
\usepackage{physics}
\usepackage{bm}

\def\mb#1{\mathbf{#1}}

\begin{document}

\title{Elasticity reshapes heat flow in graphene}

\author{Navaneetha K. Ravichandran}
\email{navaneeth@iisc.ac.in}
\affiliation{%
 Department of Mechanical Engineering, Indian Institute of Science, Bangalore 560012, India
 }%
\date{\today}
\begin{abstract}
Classical thermal transport theories that preserve rotational symmetry, predict strong anharmonic scattering of out-of-plane lattice vibrational modes called flexural phonons in flat suspended graphene sheets. Such strong scattering processes cause a breakdown of the phonon quasiparticle picture, which remains valid only when several cycles of lattice vibrations occur before the mode decays. Here we show that the renormalization of elastic bending rigidity ($D$), caused by the coupling between the in-plane and the out-of-plane thermal lattice fluctuations, restores phonon quasiparticles in suspended graphene. Importantly, this $D$-renormalization weakens the momentum-dissipating Umklapp phonon scattering processes, resulting in improved thermal conductivity and amplified phonon hydrodynamics in suspended graphene. Our results unveil a previously-unrecognized connection between the macroscopic elasticity and the microscopic flexural phonon scattering in two-dimensional (2D) materials that does not occur in three-dimensional bulk crystals, thereby motivating a re-examination of the classical theories and opening up new avenues to engineer the thermal as well as the phonon-limited electronic transport and relaxation in two- and lower-dimensional materials.
\end{abstract}
                              
\maketitle
Ever since it was first exfoliated in the monolayer form~\cite{novoselov_electric_2004}, graphene has attracted considerable attention owing to its exceptional mechanical~\cite{blees_graphene_2015}, electronic~\cite{novoselov_two-dimensional_2005} and thermal properties~\cite{seol_two-dimensional_2010, chen_thermal_2012}. Experimental research on thermal transport through graphene has been particularly active with multiple, independent experimental reports of ultrahigh thermal conductivity ($\kappa$) of suspended graphene, rivaling that of diamond, appearing within the last two decades~\cite{cai_thermal_2010, faugeras_thermal_2010, lee_thermal_2011, chen_thermal_2012, xu_length-dependent_2014, sullivan_optical_2017}. In the same time frame, the fundamental theoretical understanding of the relative strengths of different microscopic phonon scattering mechanisms, and therefore, the macroscale thermal transport properties of graphene have undergone several conceptual revisions. For example, while first-principles calculations from two decades ago involving only the lowest-order scattering among three phonons overpredicted the $\kappa$ of graphene observed experimentally at room temperature~\cite{lindsay_phonon_2014}, recent calculations including higher-order four-phonon scattering processes underpredict these experimental measurements~\cite{han_thermal_2023}. Furthermore, while previous three-phonon scattering-limited thermal transport calculations predicted strong hydrodynamic transport in suspended graphene under cryogenic conditions~\cite{lee_hydrodynamic_2015, cepellotti_phonon_2015}, recent calculations show that four-phonon scattering weakens it considerably~\cite{li_effects_2025}. In addition to these opposing predictions, there is also the fundamental question of whether the $\kappa$ of graphene converges with sample size, with the previous first-principles calculations~\cite{bonini_acoustic_2012} and experiments~\cite{xu_length-dependent_2014} reporting a logarithmically-divergent $\kappa$ with sample size, while more recent first-principles calculations involving three-phonon and four-phonon scattering processes predict a converging trend~\cite{han_thermal_2023}. Conclusive experimental confirmation of the predictions of these classical theories for heat flow in suspended monolayer graphene have been challenging due to the lack of quantitative agreement between different experimental measurements reported in the literature~\cite{cai_thermal_2010, faugeras_thermal_2010, lee_thermal_2011, chen_thermal_2012, xu_length-dependent_2014, sullivan_optical_2017}, which has been attributed to large experimental uncertainties around and below room temperature~\cite{balandin_superior_2008, lee_thermal_2011}, differences in the interpretation of raw experimental signals~\cite{balandin_superior_2008, cai_thermal_2010, lee_thermal_2011}, the possibility of non-Fourier heat flow~\cite{sullivan_optical_2017} and the effects of parasitic interfacial thermal resistances at the heaters and contacts with the substrate~\cite{jo_reexamination_2015}.

Apart from these conceptual shifts in the understanding of phonon-phonon interactions in graphene, the fundamental description of the dispersion relation of out-of-plane vibrational modes – the flexural (ZA) phonons, has also been the subject of successive reinterpretations over the past two decades. This discussion originates from the Hohenberg-Mermin-Wagner theorem~\cite{hohenberg_existence_1967, mermin_absence_1966}, which forbids long-range crystalline ordering in the flat phase of any two-dimensional (2D) system with rotational symmetry like graphene, due to thermally-driven excitations of low-energy ZA phonons with quadratic dispersions~\cite{nelson_membraneStatMech_2004}. Subsequently, the fact that large ($>$10 $\mu$m) suspended graphene monolayers were routinely being exfoliated and grown using chemical vapor deposition was reconciled by invoking a coupling between the in-plane and the out-of-plane degrees of freedom through the nonlinear terms of the elastic strain tensor~\cite{le_doussal_self-consistent_1992, aronovitz_fluctuations_1988}. The result of this reconciliation is the emergence of a renormalized bending rigidity [$D\left(L, T_0\right)$] that is no longer a material constant, but diverges with the system size, $L$, to protect the long-range ordering of the flat phase of suspended graphene sheets, thus resulting in a renormalized, sub-quadratic dispersion relation for the ZA phonons at temperatures $T_0 > 0$ K. Unlike the revised picture of phonon-phonon interactions, however, this reinterpreted theory for the ZA phonon dispersion has been confirmed experimentally through measurements of the renormalized $D$ of large suspended graphene sheets~\cite{blees_graphene_2015}.

\begin{figure*}[!ht]
    \centering
    \includegraphics[width=\linewidth, trim=10mm 10mm 20mm 2mm, clip]{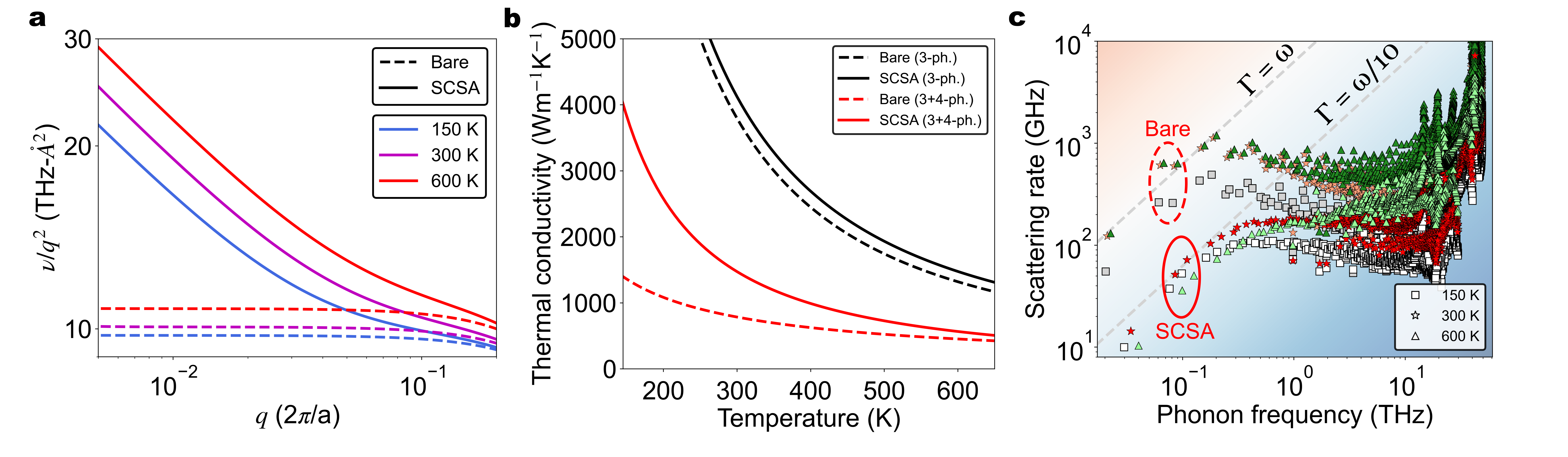}
    \caption{\textbf{SCSA renormalization of ZA phonons, and its effect on $\kappa$ and phonon quasiparticle picture in graphene.} \textbf{a}. Ratio of phonon frequency to square of the magnitude of the wave vector $\mb{q}$ ($\nu/q^2$) for the ZA phonons of graphene in the long wavelength limit. In the x-axis, $q$ is in units of $2\pi/a$, where $a$ is the lattice constant of graphene. When rotational invariance and stress-free equilibrium conditions are enforced on the harmonic IFCs in SCAP, the ZA phonons exhibit a quadratic dispersion relation. To stabilize the 2D flat phase of the suspended graphene sheet, the SCSA further renormalizes the quadratic ZA dispersion to a sub-quadratic form for $q \lesssim 0.2 (2\pi/a)$, even reaching a universal scaling of $\nu\sim q^{1.6}$  beyond a critical wavelength of $\sim$ 4 nm ($\sim$ 5 nm) at 300 K (150 K)~\cite{ravichandran_low_2026}. \textbf{b}. Calculated temperature-dependent $\kappa$ of naturally-occurring graphene with the lowest-order three-phonon scattering, $\kappa^{\left(3\right)}$, and with the additional inclusion of higher-order four-phonon scattering, $\kappa^{\left(3+4\right)}$, calculated with the bare and SCSA ZA phonons. The $\kappa^{\left(3+4\right)}$ is strongly affected by SCSA renormalization, while the $\kappa^{\left(3\right)}$ remains nearly unchanged. Comparisons with available experimental data are presented in the Supplementary Fig. S1a-b. \textbf{c}. Total phonon scattering rates ($\Gamma$) vs. frequency ($\nu$) for naturally-occurring graphene sheets at 150 K, 300 K and 600 K. The two dashed grey lines indicate $\Gamma = \omega (= 2\pi \nu)$ and $\Gamma = \omega/10$. The scattering rates for the bare ZA phonons are significantly stronger than those of the SCSA phonons over the entire Brillouin zone. In fact, $\Gamma$ exceeds $\omega$ at room temperature and beyond for the low-frequency bare ZA phonons, resulting in the breakdown of the phonon quasiparticle picture. Introduction of the SCSA renormalization restores the quasiparticle nature of these low-frequency ZA phonons with $\Gamma \lesssim \omega/10$ even up to 600 K.} \label{fig:figure_1}
\end{figure*}

These reinterpretations of thermal phonon scattering and ZA phonon dispersions have evolved relatively independently of each other, even though it is well-known that slight changes in the phonon dispersion relations can cause large differences in the phonon scattering probabilities, and consequently, the $\kappa$ of materials~\cite{ravichandran_phonon-phonon_2020}. Hence, a natural question arises at this point - does the renormalization of the macroscale elastic constant - $D$, and the resulting renormalization of the ZA phonons, affect the microscopic phonon-phonon interactions and thermal transport in graphene? Here we demonstrate, using first-principles calculations, that the renormalization of $D$ enhances the $\kappa$ and amplifies hydrodynamic phonon transport in suspended graphene sheets. These improvements occur due to the restoration of the phonon quasiparticles by the sharp weakening of the scattering processes involving the renormalized sub-quadratic ZA phonons, while the bare ZA phonons with quadratic dispersions undergo much stronger scattering that drives a complete breakdown of the quasiparticle picture. Our results uncover a previously-unexplored pathway through which macroscale elasticity controls the strength of microscopic phonon-phonon interactions, that is unique to two and lower dimensions, thus motivating a re-examination of the classical theories of heat flow in graphene and other low-dimensional systems, that did not include these effects. 

\begin{figure*}[!ht]
    \centering
    \includegraphics[width=\linewidth, trim=10mm 10mm 5mm 4mm, clip]{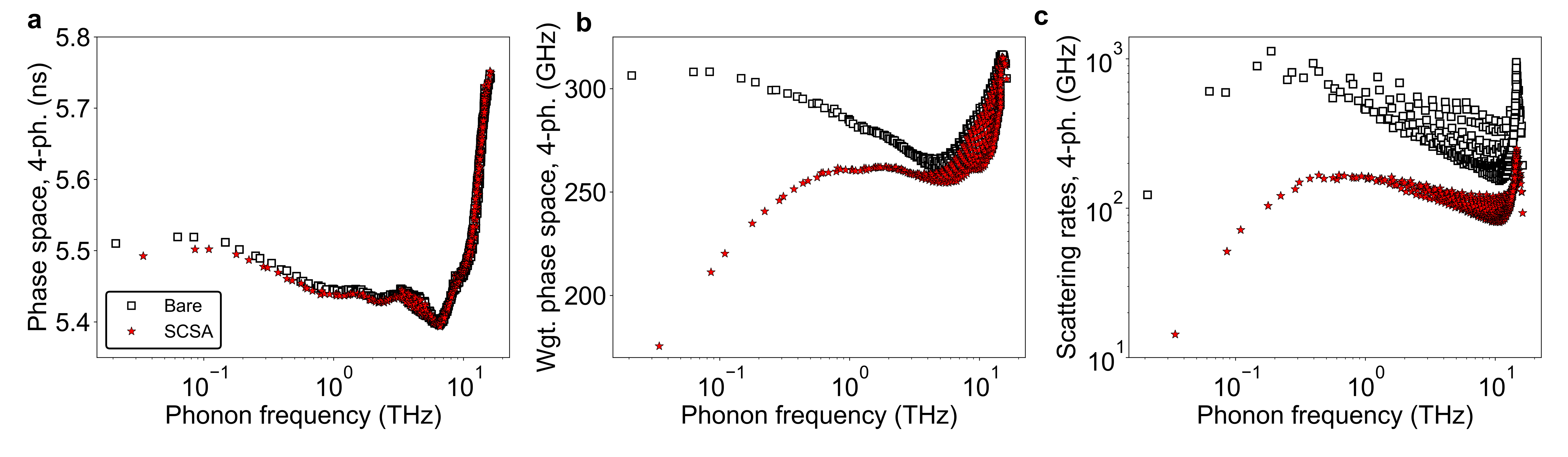}
    \caption{\textbf{Microscopic origin of weak four-phonon scattering of the SCSA ZA phonons in graphene.} \textbf{a}. Four-phonon scattering phase space, \textbf{b}. Four-phonon scattering matrix element-weighted phase space and, \textbf{c}. Four-phonon scattering rates of the ZA phonons, with and without the SCSA renormalization, at 300 K. Since the SCSA renormalization affects only the low-frequency ZA phonons, their four-phonon scattering phase space remains unaffected. However, the scattering matrix element-weighted phase space and the scattering rates for the four-phonon processes are strongly suppressed for the ZA phonons due to the SCSA renormalization, as discussed in the main text.} \label{fig:figure_2}
\end{figure*}

We compute the phonons and $\kappa$ of graphene using the self-consistent anharmonic phonon (SCAP) theory introduced in Ref.~\cite{ravichandran_unified_2018}. In this approach, the bare harmonic interatomic force constants (IFCs) are obtained from density functional perturbation theory (DFPT) and renormalized in a self-consistent manner by the anharmonic IFCs obtained using a thermal snapshot technique, while enforcing the point group symmetries, translational and rotational invariance, and stress-free equilibrium conditions. These renormalized IFCs are used to compute the phonon frequencies, eigenvectors, group velocities and phonon scattering probabilities, which are then used to obtain the $\kappa$ of graphene by solving the linearized Peierls-Boltzmann equation (LPBE) for phonon transport. Here, the scattering probabilities corresponding to three-phonon, four-phonon and phonon-isotope scattering events are included while computing the phonon collision operator, $\mb{\Omega}$. Finally, the LPBE is solved in eigenbasis of $\mb{\Omega}$ to obtain $\kappa$~\cite{ravichandran_unified_2018, ravichandran_phonon-phonon_2020, malviya_efficient_2025}, as detailed in the Methods section~\ref{sec:Methods_kappa}. To predict the role of $D$-renormalization in affecting the $\kappa$ of graphene, the ZA phonon dispersions are further renormalized using a self-consistent screening approximation (SCSA)~\cite{ravichandran_low_2026} before computing the scattering probabilities and solving the LPBE, as described in the Methods section~\ref{sec:Methods_SCSA}.

When rotational invariance and stress-free equilibrium conditions are enforced on the harmonic IFCs in SCAP, the resulting dispersion relations for the ZA phonons, which we refer to as the \emph{bare} phonons, have a quadratic dependence on the magnitude of the wave vector, $q$, as shown in Fig.~\ref{fig:figure_1}a. However, thermal fluctuations arising from these low-energy bare ZA phonons with quadratic dispersions destabilize the flat phase of suspended graphene~\cite{nelson_membraneStatMech_2004}. Upon applying the SCSA corrections to the ZA phonons that protect this flat graphene phase, their dispersions deviate from the quadratic form, and approach a limiting behavior of $\nu\sim q^{1.6}$ at small $q$~\cite{aronovitz_fluctuations_1988, le_doussal_self-consistent_1992, gazit_structure_2009, zakharchenko_self-consistent_2010}. We have shown in Ref.~\cite{ravichandran_low_2026} that this sub-quadratic ZA phonon dispersion arises out of a temperature- and system size-dependent renormalized bending rigidity [$D\left(L, T_0\right)$] in several 2D monolayers and our predictions agree well with the available experimental measurements on graphene sheets~\cite{blees_graphene_2015}.

In Fig.~\ref{fig:figure_1}b, we observe a strong influence of the SCSA corrections to the ZA phonons on the $\kappa^{\left(3+4\right)}$ of naturally-occurring graphene computed with three- and four-phonon scattering  processes. Without the SCSA corrections, the predicted $\kappa^{\left(3+4\right)}$ of naturally-occurring graphene at room temperature (300 K) is about 850 Wm$^{-1}$K$^{-1}$, while the predictions with the SCSA corrections approach 1600 Wm$^{-1}$K$^{-1}$. Furthermore, the $\kappa^{\left(3+4\right)}$ including the SCSA corrections agree reasonably well with multiple experimental measurements above 300 K, while those computed with the bare ZA phonons consistently underpredict the experiments across the entire temperature range, as shown in the Supplementary Fig. S1. We note that, in stark contrast to $\kappa^{\left(3+4\right)}$, the calculated $\kappa^{\left(3\right)}$ using only three-phonon scattering significantly overpredicts the experimental measurements (see Supplementary Fig. S1a-b), and is relatively insensitive to the SCSA corrections of the ZA phonon dispersions in Fig.~\ref{fig:figure_1}b, consistent with the observations in Ref.~\cite{lindsay_flexural_2010}, where $\kappa^{\left(3\right)}$ was computed using the ZA phonons renormalized with an approximate one-loop correction from Ref.~\cite{mariani_flexural_2008}.

The obtained low $\kappa^{\left(3+4\right)}$ of graphene using bare ZA phonons results from strong phonon-phonon scattering, which also leads to failure of the phonon quasiparticle picture. For phonons to behave as well-defined quasiparticles, their decay times must exceed several cycles of lattice vibrations, i.e., $1/\Gamma \gg 1/\omega \implies \Gamma \ll \omega$, where $\Gamma$ and $\omega$ are the scattering rates and angular frequencies of phonons, respectively. In Fig.~\ref{fig:figure_1}c, the total scattering rates are significantly stronger for the calculations using the bare ZA phonons compared to those using the SCSA ZA phonons, and even exceed $\omega$ at low frequencies, resulting in ill-defined phonon quasiparticles even at 300 K. Therefore, in retrospect, the predicted $\kappa^{\left(3+4\right)}$ from the LPBE solution, which assumes the validity of the phonon quasiparticle picture, may not be realistic when the bare ZA phonon dispersions are used. On the other hand, the SCSA corrections to the ZA phonon dispersions lower the phonon scattering rates considerably, thus restoring the phonon quasiparticle picture and admitting the analysis of heat flow in graphene within the LPBE framework. 

\begin{figure*}[!ht]
    \centering
    \includegraphics[width=\linewidth, trim=5mm 10mm 20mm 2mm, clip]{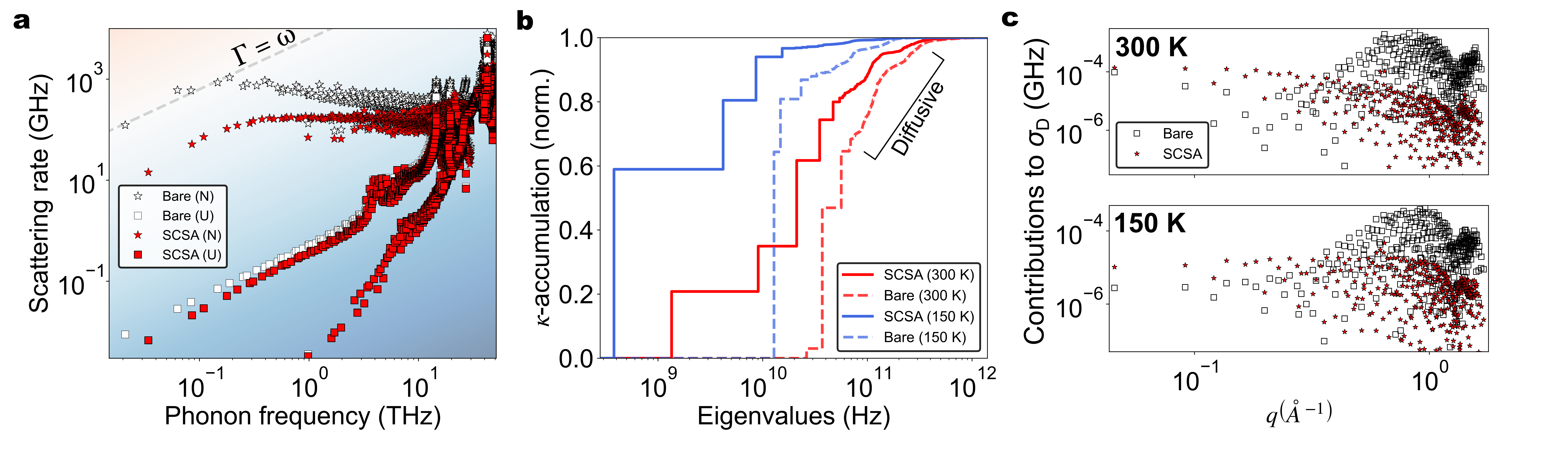}
    \caption{\textbf{Origin of the amplified phonon hydrodynamics upon SCSA renormalization of ZA phonons in graphene}. \textbf{a}. Momentum-conserving Normal (N) and momentum-dissipating Umklapp (U) scattering rates vs. $\nu$ for isotopically pure graphene at 300 K. N scattering is much stronger than U scattering for the calculations with and without SCSA renormalization of the ZA phonons, even at 300 K. The difference between N and U scattering rates is greater for the calculations with the bare ZA phonons than for those with the SCSA ZA phonons. Yet, $\kappa^{\left(3+4\right)}_{\text{SCSA}} > \kappa^{\left(3+4\right)}_{\text{bare}}$, contrary to conventional understanding. \textbf{b}. Cumulative $\kappa$ contributions from the eigenmodes of the phonon collision operator ($\mb{\Omega}$) in isotopically pure graphene at 150 K and 300 K for the calculations with the bare and the SCSA ZA phonons. The $\kappa$-accumulation computed with the SCSA ZA phonons shows large contribution to $\kappa^{\left(3+4\right)}$ from individual eigenmodes with smaller eigenvalues (called the drift eigenmodes, $\sigma^D$, in the main text) compared to that computed with the bare ZA phonons at both temperatures, indicating amplified hydrodynamic heat flow due to SCSA renormalization of the ZA phonons. \textbf{c}. Contributions of phonons to $\sigma^D$ at 150 K and 300 K in graphene, with and without the SCSA renormalization of the ZA phonons. $\sigma_D \approx \mel{\mathfrak{e}^D}{\mb{\Omega}^{\left(U\right)}}{\mathfrak{e}^D}/\braket{\mathfrak{e}^D}{\mathfrak{e}^D}$ contains contributions of individual phonons to the matrix element of the collision operator for U processes only, which are smaller for the calculations using the SCSA ZA phonons compared to those using the bare ZA phonons, thus lowering $\sigma^D$ and amplifying phonon hydrodynamics in the former.} \label{fig:figure_3}
\end{figure*}

At this point, we distinguish our findings from those of Ref.~\cite{aseginolaza_bending_2024}, where the classical harmonic theory with quadratic ZA phonon dispersions was shown to cause the breakdown of the quasiparticle picture for sound waves, i.e., the long-wavelength in-plane longitudinal (LA) and transverse (TA) acoustic phonons, which, however, gets restored by anharmonic renormalization using the lowest-order bubble interaction diagram, without affecting the quadratic dispersion of the ZA phonons. In contrast, here we show that the higher-order four-phonon interactions among the bare ZA phonons with quadratic dispersions that go beyond the bubble interaction diagram, cause the breakdown of the ZA quasiparticles themselves, thus necessitating their renormalization through the SCSA framework to restore their quasiparticle nature. This SCSA renormalization of the ZA phonons also restores the quasiparticle nature of the LA and TA thermal phonons, as shown in Supplementary Fig. S2. 

We further investigate, in Fig.~\ref{fig:figure_2}, the origin of the weaker phonon scattering in graphene with the SCSA ZA phonons. The SCSA corrections predominantly affect the low frequency, long wavelength ZA phonons owing to the ease of their thermal excitation, as evident from Fig.~\ref{fig:figure_1}a for phonons with wavelengths longer than $\sim$ 1.5 nm [or $q \lesssim 0.16\ (2\pi/a)$, where $a$ is the lattice constant]. Hence, these frequency corrections do not affect their scattering phase space from three-phonon (Supplementary Fig. S3a) and four-phonon processes (Fig.~\ref{fig:figure_2}a), since they only appear in summations with the higher frequencies of other phonons, within the arguments of the Dirac delta functions for phase space (see Methods section~\ref{sec:Methods_kappa}). On the other hand, Fig.~\ref{fig:figure_2}b shows that the SCSA renormalization strongly suppresses the matrix element-weighted four-phonon scattering phase space ($\chi_{4-ph.}$) of the low frequency ZA phonons - a trend that arises from the weaker four-phonon matrix elements for the SCSA phonons, as elucidated in the Supplementary Note 1. Our calculations also recover the expected frequency independence of $\chi_{4-ph.}$ for the low-frequency bare ZA phonons with quadratic dispersions in Fig.~\ref{fig:figure_2}b, as derived in the Supplementary Note 1. The four-phonon scattering rates ($\Gamma_{4-ph.}$) result from further weighting $\chi_{4-ph.}$ with Bose factors, which are much smaller at a given $q$ for the SCSA ZA phonons owing to their higher frequencies compared to the bare ZA phonons. Hence, rather than being a phase space effect, the weaker $\Gamma_{4-ph.}$ for the SCSA ZA phonons relative to their bare counterparts observed in Fig.~\ref{fig:figure_2}c arises from weaker four-phonon scattering matrix elements and smaller Bose factors for the former. The frequency dependence of three-phonon scattering rates ($\Gamma_{3-ph.}$) for the ZA phonons in the Supplementary Fig. S3c match with those reported in Refs.~\cite{bonini_acoustic_2012, lindsay_phonon_2014}, and the relatively smaller differences in $\Gamma_{3-ph.}$ for the sub-quadratic SCSA and the quadratic bare ZA phonons in Supplementary Fig. 3c are also consistent with the reported insensitivity of $\Gamma_{3-ph.}$ in Ref.~\cite{lindsay_flexural_2010} to the approximate renormalization of the ZA phonons following Ref.~\cite{mariani_flexural_2008}. 

Interestingly, the momentum-conserving Normal (N) scattering rates for the calculations using the bare ZA phonons are much stronger than those using the SCSA ZA phonons at 300 K (Fig.~\ref{fig:figure_3}a), even though the $\kappa$ for the former is much lower ($\approx$ 50\%) than that of the latter (Fig.~\ref{fig:figure_1}b). This unexpected trend originates from the large difference between the N and the momentum-dissipating Umklapp (U) scattering strengths in the calculations using the bare as well as the SCSA ZA phonons, as observed for the N and the U scattering rates in Fig.~\ref{fig:figure_3}a. In the ideal limit of vanishingly small U processes, i.e., $\mb{\Omega}^{\left(N\right)} \gg \mb{\Omega}^{\left(U\right)}$, we have shown in the Supplementary Note 2 that the $\kappa = \frac{\left|\mel{\mathfrak{e}^D}{\mb{v}_\lambda}{\mathfrak{e}^0}\right|^2}{\sigma_D\braket{\mathfrak{e}^D}{\mathfrak{e}^D}}$, where $\ket{\mathfrak{e}^0}$ and $\ket{\mathfrak{e}^D}$ are the equilibrium and drift eigenmodes of the phonon collision operator $\mb{\Omega} = \mb{\Omega}^{\left(N\right)} + \mb{\Omega}^{\left(U\right)} \approx \mb{\Omega}^{\left(N\right)}$ with eigenvalues $0$ and $\sigma_D = \frac{\mel{\mathfrak{e}^D}{\mb{\Omega}}{\mathfrak{e}^D}}{\braket{\mathfrak{e}^D}{\mathfrak{e}^D}}$ respectively. In this ideal limit, which occurs under cryogenic conditions, the entire contribution to $\kappa$ is from the drift eigenmodes only. 

Although this ideal limit does not apply directly at 150 K and 300 K, the $\kappa$-contributions of the eigenmodes of $\mb{\Omega}$ in Fig.~\ref{fig:figure_3}b reveal that the calculations using the SCSA and the bare ZA phonons exhibit large drifting components, identified by large $\kappa$-contributions from individual drift eigenmodes with small eigenvalues ($\sigma^D$), but much smaller diffusive, non-drifting contributions from larger eigenvalues, at 150 K and 300 K. However, the $\sigma^D$ for the calculations with the SCSA ZA phonons are much smaller than those with the bare ZA phonons. Since in the limit of vanishing U processes - the ideal limit of pure hydrodynamic heat flow, the drift modes being the null vectors of $\mb{\Omega} \approx \mb{\Omega}^{\left(N\right)}$ have vanishing $\sigma^D$ (see Methods section~\ref{sec:Methods_SCSA} and Supplementary Note 2), the observed smaller $\sigma^D$ from the calculations with the SCSA ZA phonons in Fig.~\ref{fig:figure_3}b indicates amplified phonon hydrodynamics relative to those with the bare ZA phonons. 

The smaller $\sigma^D$ in the calculations with the SCSA ZA phonons arises from the weaker U processes that they undergo relative to the bare ZA phonons. In Fig.~\ref{fig:figure_3}c, large-$q$ SCSA ZA phonons contribute less to individual elements of $\sigma^D = \frac{\mel{\mathfrak{e}^D}{\mb{\Omega}}{\mathfrak{e}^D}}{\braket{\mathfrak{e}^D}{\mathfrak{e}^D}}$ relative to their bare ZA counterparts, at both 150 K and 300 K. Since these contributions arise entirely out of the matrix elements of $\mb{\Omega}^{\left(U\right)}$ (since $\mel{\mathfrak{e}^D}{\mb{\Omega}^{\left(N\right)}}{\mathfrak{e}^D} = 0$), the weaker U scattering of the SCSA ZA phonons relative to the bare ZA phonons (i.e., $\mb{\Omega}^{\left(U\right)}_{\text{SCSA}} < \mb{\Omega}^{\left(U\right)}_{\text{bare}}$) results in a strongly amplified hydrodynamic drifting contribution to $\kappa$ in the former.

Our results, thus, uncover a previously-unrecognized fundamental role of the macroscale elasticity in controlling the microscopic phonon-phonon interactions, thereby preserving phonon quasiparticles, improving heat conduction and amplifying phonon hydrodynamics in suspended graphene sheets, apart from stabilizing their flat crystalline 2D phase. Our findings, however, have implications beyond heat conduction, extending to systems beyond graphene. Phonon-phonon scattering channels control the thermalization of hot electrons and spins in 2D semiconductors, thus affecting their electronic mobility, electronic noise spectra, and spin-coherence times. The conceptual reformulation of phonons and their decay mechanisms in 2D systems introduced in this work opens up new avenues for engineering novel thermal, electronic and quantum phenomena in these materials, that are difficult to realize in bulk three-dimensional systems. By virtue of the generality of the first-principles thermal transport framework introduced here, our work will also accelerate the search for unconventional, non-Fourier heat flow regimes in other 2D as well as lower-dimensional systems, whose heat-carrying flexural vibrations are sensitive to the thermally-amplified nonlinearities in the elastic properties. 

This work was supported by the Core Research Grant (CRG) No. CRG/2022/009160, and the Mathematical Research Impact Centric Support (MATRICS) Grant No. MTR/2022/001043 from the Science and Engineering Research Board, India, and by the Advanced Research Grant (ARG) No. ANRF/ARG/2025/007160/ENS from the Anusandhan National Research Foundation, India. NR thanks David Broido and Nikhil Malviya for useful discussions.

\section*{Methods}
\subsection{First-principles calculation of thermal conductivity}   \label{sec:Methods_kappa}
The thermal conductivity, $\kappa$, of graphene is obtained by solving the linearized Peierls-Boltzmann equation for phonon transport under steady state conditions, given by:
\begin{align}
    \mb{v}_\lambda\cdot\nabla T\frac{\partial n^0_\lambda}{\partial T} = \mathcal{C}\left(\tilde{n}_{\lambda}\right)  \label{eq:SSPBE}
\end{align}
where $\mb{v}_\lambda$ and $n^0_\lambda$ are the group velocities and the equilibrium Bose-Einstein distribution functions of the phonon mode $\lambda\equiv\left(\mb{q}, j\right)$ with wave vector $\mb{q}$ and polarization $j$, $\tilde{n}_\lambda = \frac{n_\lambda - n^0_\lambda}{n^0_\lambda\left(n^0_\lambda + 1\right)}$ is the deviational non-equilibrium distribution function derived from the total non-equilibrium distribution, $n_\lambda$, and $\mathcal{C}\left(\tilde{n}_{\lambda}\right) = \mathcal{C}_{3}\left(\tilde{n}_{\lambda}\right) + \mathcal{C}_{4}\left(\tilde{n}_{\lambda}\right) + \mathcal{C}_{iso.}\left(\tilde{n}_{\lambda}\right)$ is the linearized phonon collision integral describing three-phonon, four-phonon and phonon-isotope scattering processes. The collision integral for different phonon scattering processes are given by:
\begin{align}
    \mathcal{C}_{3}&\left(\tilde{n}_{\lambda}\right) = \sum_{\lambda_1\lambda_2}\Bigg[\mathcal{W}^{+}_{\lambda\lambda_1\lambda_2}\left(\tilde{n}_\lambda + \tilde{n}_{\lambda_1} - \tilde{n}_{\lambda_2}\right) \nonumber\\
    &\ \ \ \ \ \ \ \ + \frac{1}{2}\mathcal{W}^{-}_{\lambda\lambda_1\lambda_2}\left(\tilde{n}_\lambda - \tilde{n}_{\lambda_1} - \tilde{n}_{\lambda_2}\right)\Bigg]\nonumber\\
    \mathcal{C}_{4}&\left(\tilde{n}_{\lambda}\right) = \sum_{\lambda_1\lambda_2\lambda_3}\Bigg[\frac{1}{2}\mathcal{Y}^{++}_{\lambda\lambda_1\lambda_2\lambda_3}\left(\tilde{n}_\lambda + \tilde{n}_{\lambda_1} + \tilde{n}_{\lambda_2} - \tilde{n}_{\lambda_3}\right) \nonumber\\
    &\ \ \ \ \ \ \ + \frac{1}{2}\mathcal{Y}^{+-}_{\lambda\lambda_1\lambda_2\lambda_3}\left(\tilde{n}_\lambda + \tilde{n}_{\lambda_1} - \tilde{n}_{\lambda_2} - \tilde{n}_{\lambda_3}\right) \nonumber\\
    &\ \ \ \ \ \ \ + \frac{1}{6}\mathcal{Y}^{--}_{\lambda\lambda_1\lambda_2\lambda_3}\left(\tilde{n}_\lambda - \tilde{n}_{\lambda_1} - \tilde{n}_{\lambda_2} - \tilde{n}_{\lambda_3}\right)\Bigg]\nonumber\\
    \mathcal{C}_{iso.}&\left(\tilde{n}_{\lambda}\right) = \sum_{\lambda_1}\mathcal{X}_{\lambda\lambda_1}\left(\tilde{n}_\lambda - \tilde{n}_{\lambda_1}\right)  \label{eq:coll_integrals}
\end{align}
where the scattering probabilities are given by,
\begin{align}
    \mathcal{W}^{\pm}_{\lambda\lambda_1\lambda_2} &= -\frac{\pi\hbar}{4N_0}\left|\Phi_{\lambda\left(\pm\lambda_1\right)\left(-\lambda_2\right)}\right|^2\delta\left(\omega_{\lambda} \pm \omega_{\lambda_1} - \omega_{\lambda_2}\right)\nonumber\\
    &\ \ \ \ \ \ \ \ \times n^0_\lambda\left(n^0_{\lambda_1}+\frac{1}{2}\mp\frac{1}{2}\right)\left(n^0_{\lambda_2} + 1\right)\nonumber
\end{align}
\begin{align}
    \mathcal{Y}^{\pm\pm}_{\lambda\lambda_1\lambda_2\lambda_3} &= -\frac{\pi\hbar}{4N_0}\left|\Phi_{\lambda\left(\pm\lambda_1\right)\left(\pm\lambda_2\right)\left(-\lambda_3\right)}\right|^2\nonumber\\
    &\ \ \ \ \ \ \times\delta\left(\omega_{\lambda} \pm \omega_{\lambda_1} \pm \omega_{\lambda_2} - \omega_{\lambda_3}\right)n^0_\lambda\left(n^0_{\lambda_3} + 1\right)\nonumber\\
    &\ \ \ \ \ \ \times \left(n^0_{\lambda_1}+\frac{1}{2}\mp\frac{1}{2}\right)\left(n^0_{\lambda_2}+\frac{1}{2}\mp\frac{1}{2}\right)\nonumber
\end{align}
\begin{align}
    \mathcal{X}_{\lambda\lambda_1} = &-\frac{\pi\omega_\lambda^2}{2N_0}\left|\Psi_{\lambda\left(-\lambda_1\right)}\right|^2\delta\left(\omega_\lambda - \omega_{\lambda_1}\right)n^0_\lambda\left(n^0_\lambda + 1\right)    \label{eq:scat_prp}
\end{align}
with,
\begin{align}
    \Phi_{\lambda\lambda_1\lambda_2} =& \sum_{l_1l_2}\sum_{kk_1k_2}\sum_{\alpha\beta\gamma}\frac{\Phi_{\alpha\beta\gamma}^{\left(0k, l_1k_1, l_2k_2\right)}}{\sqrt{m_km_{k_1}m_{k_2}}}\frac{e^{i\left(\mb{q}_1\cdot\mb{R}_{l_1} + \mb{q}_2\cdot\mb{R}_{l_2}\right)}}{\sqrt{\omega_\lambda\omega_{\lambda_1}\omega_{\lambda_2}}}\nonumber\\
    \times w_\alpha&\left(\lambda, k\right)w_\beta\left(\lambda_1, k_1\right)w_\beta\left(\lambda_2, k_2\right)\Delta\left(\mb{q} + \mb{q}_1 + \mb{q}_2\right)\nonumber
\end{align}
\begin{align}
    \Phi_{\lambda\lambda_1\lambda_2\lambda_2} &= \sum_{l_1l_2l_3}\sum_{kk_1k_2k_3}\sum_{\alpha\beta\gamma\delta}\frac{\Phi_{\alpha\beta\gamma\delta}^{\left(0k, l_1k_1, l_2k_2, l_3k_3\right)}}{\sqrt{m_km_{k_1}m_{k_2}m_{k_3}}}\nonumber\\
    \times e&^{i\left(\mb{q}_1\cdot\mb{R}_{l_1} + \mb{q}_2\cdot\mb{R}_{l_2} + \mb{q}_2\cdot\mb{R}_{l_3}\right)}\frac{\Delta\left(\mb{q} + \mb{q}_1 + \mb{q}_2 + \mb{q}_3\right)}{\sqrt{\omega_\lambda\omega_{\lambda_1}\omega_{\lambda_2}\omega_{\lambda_3}}}\nonumber\\
    \times w_\alpha&\left(\lambda, k\right)w_\beta\left(\lambda_1, k_1\right)w_\beta\left(\lambda_2, k_2\right)w_\gamma\left(\lambda_3, k_3\right)\nonumber
\end{align}
\begin{align}
    \left|\Psi_{\lambda\lambda_1}\right|^2 = &\sum_{ik}\left[f_i\left(k\right)\left(1-\frac{m_{k, i}}{\bar{m}_k}\right)^2\right]\left|\mb{w}\left(\lambda, k\right)\cdot\mb{w}\left(\lambda_1, k\right)\right|^2
\end{align}
Here, $l_i$ are the lattice sites with lattice vectors $\mb{R}_{l_i}$, $k$ and $k_i$ are the indices for the basis atoms, $\mb{w}\left(\lambda, k\right)$ are the eigenvectors for the phonon mode $\lambda$ and the basis atom $k$, $\left(\alpha, \beta, \gamma, \delta\right)$ are the Cartesian indices, $m_{k, i}$ and $f_i\left(k\right)$ are the mass and the concentration of the $i^{th}$ isotope of the atom at the basis site $k$ respectively, with $\bar{m}_k$ being the isotopic average mass of the atom at the basis site $k$, $\Phi_{\alpha\beta\gamma}^{\left(0k, l_1k_1, l_2k_2\right)}$ and $\Phi_{\alpha\beta\gamma\delta}^{\left(0k, l_1k_1, l_2k_2, l_3k_3\right)}$ are the components of the cubic and the quartic interatomic force constants (IFCs) respectively and $N_0$ is the number of unit cells in the crystal. Here, $\Delta\left(\cdot\right)$ represents the Kronecker delta function, which results in a value of $1$ if the argument is $0$ modulo any reciprocal lattice vector $\mb{G}$, and $0$ otherwise. For a set of phonons participating in a scattering process, if $\mb{G}=0$, then the process is momentum-conserving [Normal (N) process]. Otherwise, it is a momentum-dissipating [Umklapp (U)] process. We note that the phonon-isotope scattering processes are momentum-dissipative by nature, and are, therefore, included in the U processes in the main text for naturally-occurring materials. The coefficients of $\tilde{n}_\lambda$ in $\mathcal{C}_3\left(\tilde{n}_\lambda\right)$, $\mathcal{C}_4\left(\tilde{n}_\lambda\right)$ and $\mathcal{C}_{iso.}\left(\tilde{n}_\lambda\right)$ in Eq.~\ref{eq:coll_integrals} are $n^0_\lambda\left(n^0_\lambda + 1\right)\Gamma_{3-ph.}$, $n^0_\lambda\left(n^0_\lambda + 1\right)\Gamma_{4-ph.}$ and $n^0_\lambda\left(n^0_\lambda + 1\right)\Gamma_{iso.}$ respectively, where $\Gamma$'s are the respective scattering rates. The scattering matrix element-weighted phase space (Fig.~\ref{fig:figure_2}b) is obtained by evaluating $\Gamma$ without the Bose factors in the scattering probabilities (Eq.~\ref{eq:scat_prp}), and the phase space (Fig.~\ref{fig:figure_2}a) is obtained by eliminating all but the Dirac-delta functions and the constant factors.

We obtain the phonon frequencies, eigenvectors and group velocities first from density functional perturbation theory (DFPT) using Quantum ESPRESSO (QE)~\cite{giannozzi_quantum_2009}. Subsequently, we obtain the cubic and the quartic IFCs using the thermal snapshot technique~\cite{ravichandran_unified_2018} with the necessary force-displacement dataset obtained from density functional theory (DFT) as implemented in QE. These anharmonic IFCs are finally used to obtain the thermally-renormalized phonons as well as the thermally-renormalized cubic and quartic IFCs in a self-consistent manner using the self-consistent anharmonic phonon (SCAP) framework~\cite{ravichandran_unified_2018}. We refer to these thermally-renormalized phonons obtained from the SCAP as the \emph{bare} phonons in the main text (shown in Ref.~\cite{ravichandran_low_2026} along the special symmetry directions), to distinguish from the additional renormalization of the ZA phonons that is necessary to stabilize 2D systems. For the DFPT calculations in QE, we obtained a convergence of $< 10^{-3}$ Ry for the total energy, $< 0.3$ kbar per unit cell for the total stress and $< 10^{-5}$ Ry/au for the forces, by using a kinetic energy cutoff of 105 Ry for the wave function, 420 Ry for the electronic density, a $\Gamma$-shifted 30$\times$30$\times$1 electronic $\mb{k}$-grid with a Marzari-Vanderbilt smearing of 0.02 Ry and a 9$\times$9$\times$1 $\Gamma$-centered phonon $\mb{q}$-grid. For the DFT calculations of the force-displacement dataset on the thermal snapshots, we used the same energy cut-offs as in the DFPT calculations and performed $\Gamma$-centered DFT calculations on 200 thermal snapshots to efficiently sample the anharmonic IFCs, as discussed in Ref.~\cite{ravichandran_unified_2018}.

To solve the linear dynamical system in Eq.~\ref{eq:SSPBE}, we rewrite Eq.~\ref{eq:SSPBE} following Ref.~\cite{cepellotti_thermal_2016, malviya_efficient_2025} as:
\begin{align}
    \frac{\mb{v}_\lambda\cdot\nabla T}{\sqrt{n^0_\lambda\left(n^0_\lambda + 1\right)}} \frac{\partial n^0_\lambda}{\partial T} = \frac{\mathcal{C}\left(\tilde{n}_\lambda\right)}{\sqrt{n^0_\lambda\left(n^0_\lambda + 1\right)}} = -\sum_{\lambda\lambda_1}\Omega_{\lambda\lambda_1}\xi_{\lambda_1}   \label{eq:SSPBE_eig}
\end{align}
where $\xi_{\lambda_1} = \sqrt{n^0_{\lambda_1}\left(n^0_{\lambda_1} + 1\right)}\tilde{n}_{\lambda_1}$ and $\mb{\Omega}\equiv\Omega_{\lambda\lambda_1}$ is the collision operator that has been made symmetric in $\left(\lambda, \lambda_1\right)$. Since the eigenvectors of the symmetric matrix $\mb{\Omega}$, given by $\{\ket{\mathfrak{e}_i}\}$, form a complete orthonormal basis, the solution to Eq.~\ref{eq:SSPBE_eig} can be written as $\xi_\lambda = \sum_{i}a_i\braket{\lambda}{\mathfrak{e}_i}$, where $\braket{\lambda}{\mathfrak{e}_i}$ is the representation of $\ket{\mathfrak{e}_i}$ in the phonon basis. Using this eigenmode expansion, the thermal conductivity tensor $\kappa_{\alpha\beta}$ takes the form: $\kappa_{\alpha\beta} = C_0\sum_i\frac{\mathcal{V}^{0i}_\alpha}{\sqrt{\sigma_i}}\frac{\mathcal{V}^{0i}_\beta}{\sqrt{\sigma_i}}$, where $C_0$ is the total volumetric heat capacity of the phonons, and $\mathcal{V}^{0i}_\alpha = \mel{\mathfrak{e}_0}{v_{\lambda, \alpha}}{\mathfrak{e}_i}$ is the velocity of the eigenmode $i$. Here, $\ket{\mathfrak{e}_0}$ is the eigenmode of $\mb{\Omega}$ corresponding to the equilibrium distribution; therefore, it is a null vector of $\mb{\Omega}$ since collisions do not modify an equilibrium distribution of phonons, and is given by $\braket{\lambda}{\mathfrak{e}^0} = \frac{\sqrt{n^0_\lambda\left(n^0_\lambda + 1\right)}\hbar\omega_\lambda}{\sqrt{Vk_BT_0^2C_0}}$~\cite{malviya_efficient_2025}. We checked for the convergence of $\kappa$ with respect to the discretization density of the Brillouin zone (BZ), while including three-phonon, four-phonon and phonon-isotope interactions. We find that a $65^2-75^2$ discretization of the first BZ is sufficient to achieve convergence of $\kappa$ at all temperatures considered in this work (see Supplementary Fig. S1c).

When the N processes are much stronger than the U processes, $\mb{\Omega}\approx\mb{\Omega}^{\left(N\right)}$. In this limit, $\mathcal{C}_3^{\left(N\right)}\left(\tilde{n}_\lambda\right)$ and $\mathcal{C}_4^{\left(N\right)}\left(\tilde{n}_\lambda\right)$ vanish identically when $\tilde{n}_\lambda = \tilde{n}^D_\lambda \propto \mb{q}\cdot\mb{c}$, where $\mb{c}$ is a constant vector. The corresponding non-equilibrium distribution function in Eq.~\ref{eq:SSPBE_eig} forms an additional two (three) mutually-orthogonal null vectors for $\mb{\Omega}$ in 2D (3D), corresponding to the two (three) Cartesian components of $\mb{c}$. These null vectors represent a drifting motion of the phonon gas as a whole with a velocity $\mb{c}$, since $\tilde{n}^D_\lambda \propto \mb{q}\cdot\mb{c}$ corresponds to the deviational distribution of $n^0_\lambda\left(\omega_\lambda - \mb{q}\cdot\mb{c}\right) \approx n^0_\lambda\left(\omega\right) - \mb{q}\cdot\mb{c}\frac{\partial n^0_\lambda\left(\omega\right)}{\partial\omega}$ - a shifted equilibrium distribution function. Following the convention for $\braket{\lambda}{\mathfrak{e}^0}$, the eigenvectors of $\mb{\Omega}$ corresponding to $\tilde{n}^D_\lambda$ are given by $\braket{\lambda}{\mathfrak{e}^D} = \frac{\sqrt{n^0_\lambda\left(n^0_\lambda + 1\right)}\hbar\mb{q}\cdot\mb{c}}{\sqrt{Vk_BT_0^2C_0}}$ up to a normalization constant.

\subsection{Renormalization of ZA phonons using self-consistent screening approximation}    \label{sec:Methods_SCSA}
As discussed in Ref.~\cite{ravichandran_low_2026}, the point-group symmetries, the translational and rotational invariance and the stress-free equilibrium conditions are rigorously enforced while computing the phonons as well as the anharmonic IFCs within the SCAP framework. For a rotationally-invariant suspended graphene sheet in equilibrium, the free energy takes the form: $\mathcal{F}\left(H\right) = \frac{1}{2A_0}\int D\left(\nabla^2 H\right)^2 \mathrm{d}\mb{x}$, where $H$ is the displacement in the out-of-plane direction, $A_0$ is the area of the graphene sheet and $D$ is the bare bending rigidity of the graphene sheet. In such a rotationally-invariant form of $\mathcal{F}\left(H\right)$, the in-plane and the out-of-plane degrees of freedom are decoupled, resulting in a ZA phonon dispersion with a quadratic dependence on the phonon wave vector $\mb{q}$ at small $\mb{q}$. The thermal excitation of these low energy ZA phonons with a quadratic dispersion relation destabilizes the flat phase of large 2D suspended graphene sheets, as discussed in Ref.~\cite{nelson_membraneStatMech_2004}. The stability of large suspended graphene sheets is conventionally restored by accounting for the nonlinear terms of the elastic strain tensor that couples the in-plane and the out-of-plane degrees of freedom, resulting in a renormalized bending rigidity $D\left(L, T_0\right)$ that depends on the system size $L\sim 2\pi/\|\mb{q}\|_2$ and temperature $T_0$, and consequently, a renormalized sub-quadratic ZA phonon dispersion~\cite{aronovitz_fluctuations_1988, le_doussal_self-consistent_1992, ravichandran_low_2026}.

In this work, we employ the self-consistent screening approximation (SCSA) with first-principles inputs, as implemented in our recent work~\cite{ravichandran_low_2026}, to renormalize the bending rigidity and the ZA phonon dispersions of 2D monolayers. The renormalized $D\left(\mb{q}, T_0\right)$ is obtained by solving the Dyson's equation for the renormalized propagator [$G^{-1}\left(\mb{q}, T_0\right)$]:
\begin{align}
    \frac{D\left(\mb{q}, T_0\right)}{k_BT_0} = G^{-1}\left(\mb{q}, T_0\right) = G_0^{-1}\left(\mb{q}, T_0\right) + \Pi\left(\mb{q}, T_0\right)
\end{align}
in a self-consistent manner. Here, $G_0^{-1}\left(\mb{q}, T_0\right) = \frac{D\left(T_0\right)}{k_BT_0}q^4$ is the bare propagator and the self energy, $\Pi\left(\mb{q}, T_0\right)$, is given by:
\begin{align}
    \Pi\left(\mb{q}, T_0\right) &= \int\frac{d^2\mb{k}}{2\pi^2}\frac{B_0\left(T_0\right)}{1 + B_0\left(T_0\right)I\left(\mb{k}\right)}\|\hat{\mb{k}}\times\mb{q}\|_2^4G\left(\mb{q}-\mb{k}\right)
\end{align}
where $B_0\left(T_0\right) = Y^{\text{2D}}_0\left(T_0\right)/\left(2k_B T_0\right)$, with $D\left(T_0\right)$ and and $Y^{\text{2D}}_0\left(T_0\right)$ being the bare bending rigidity and 2D Young's modulus of graphene obtained from the SCAP first-principles framework described earlier, and $I\left(\mb{k}\right) = \int\frac{\mb{d}\mb{p}}{\left(2\pi\right)^2}\left[\mb{p}P^T\left(\mb{k}\right)\mb{p}\right]^2G\left(\mb{p}\right)G\left(\mb{k}-\mb{p}\right)$ is related to the vacuum polarization integral introduced in Ref.~\cite{le_doussal_self-consistent_1992}. The renormalized ZA phonon frequencies are then obtained as $\omega_{\mathrm{ZA}} = \sqrt{D\left(\mb{q}, T_0\right)/\rho_{\mathrm{2D}}}\|\mb{q}\|^2$, where $\rho_{\mathrm{2D}}$ is the 2D mass density of the graphene sheet.

\bibliographystyle{unsrt}
\bibliography{references_NatPhys}
\end{document}